


\long\def\UN#1{$\underline{{\vphantom{\hbox{#1}}}\smash{\hbox{#1}}}$}
\def\NL{\hfill\break}
\def\NI{\noindent}
\magnification=\magstep 1
\overfullrule=0pt
\hfuzz=16pt
\voffset=0.1 true in
\vsize=9.2 true in
   \def\NP{\vfil\eject}
   \baselineskip 20pt
   \parskip 6pt
   \hoffset=0.1 true in
   \hsize=6.2 true in
\nopagenumbers
\pageno=1
\footline={\hfil -- {\folio} -- \hfil}

\hphantom{AA}

\hphantom{AA}

\hphantom{AA}

\centerline {\bf MORPHOLOGY OF AMORPHOUS LAYERS BALLISTICALLY}

\centerline {\bf DEPOSITED ON A PLANAR SUBSTRATE}

\vskip 0.4in

\centerline{\bf B.D.~Lubachevsky,$
^{a}$\ \ V.~Privman$^{b,c}$\ \ {\rm and}\ \ S.C.~Roy$^{d}$}

\vskip 0.2in

\item{$^a$}{\sl{AT\&T Bell Laboratories,
600 Mountain Avenue, Murray Hill, NJ 07974, USA}}

\item{$^b$}{\sl{Department of Physics, Theoretical Physics, University
of Oxford, 1 Keble Road, Oxford OX1 3NP, UK}}

\item{$^c$}on leave of absence from {\sl{Department of Physics,
Clarkson University, Potsdam, NY 13699, USA}}

\item{$^d$}{\sl{Department of Computer Science, College of William and
Mary, Williamsburg, VA 23185, USA}}

\vskip 0.6in

\NI {\bf PACS:}$\;$ 05.90.+m, 68.55.Jk, 82.70.Dd.

\NP

\centerline{\bf ABSTRACT}

We report numerical simulation of the deposition of
spherical particles on a planar surface,
by ballistic, straight-line trajectory transport, and assuming
irreversible adhesion on contact with the surface or previously
deposited particles. Our data indicate that the deposit formed has a
loosely layered structure within few diameters from the surface. This
structure can be explained by a model of growth via chain formation.
Away from the surface we found evidence of a monotonic,
power-law approach to the bulk density. Both density and
contact-statistics
results suggest that the deposit formed is sparse: the space-filling
fraction is about 15\%, and the average number of
contacts is 2. The morphology of the deposit both near the surface
and in the bulk seems to be a result of competition of screening
and branching; nearly half of all the spheres are either single-contact
dangling ends, or branching nodes with more than two contacts.

\NP

\NI{\bf 1.~INTRODUCTION}

\hphantom{AA}

Formation of deposits of proteins and monodisperse colloid particles
on planar substrates (surfaces)
has been a topic of recent theoretical interest
[1] due to advances in experimental realizations and characterizations
of such systems [2]. Since analytical theories are at best approximate
[1], formation of multilayer deposits has been studied
mostly by numerical means [1]. However, these numerical studies were
limited to lattice models, and many of the detailed results were
obtained only for the deposition on linear, one-dimensional
``substrates.'' In this work we report the first detailed numerical
investigation of the morphology of off-lattice deposits of spherical
particles at a planar, two-dimensional wall. We also introduce a new
theoretical model which explains semi-quantitatively the numerical
findings.

Our interest is mainly in the morphology of the deposit near the wall
(i.e., the structure of the first few layers).
Indeed, deposition experiments [1,2] providing motivation for our study
typically involve formation of not too many layers. (Note that the
concept of ``layers'' is at best approximate in random,
amorphous structures.) As emphasized in earlier studies [1],
the transport
mechanism of particles to the wall is then not a decisive feature. Here
the simplest, ballistic-deposition transport will be assumed: spherical
particles are dropped randomly over the substrate. They fall along
straight-line trajectories (perpendicular to the substrate) and stick
irreversibly, at the first contact, to the substrate or to one of
the particles deposited earlier.

We only report results for the final-state structure after many spheres
were deposited and the near-wall layers reached saturation. Study of
the time-dependent properties is also of interest. However, all such
investigations are rather computer-resource (time and memory) demanding.
Ballistic deposition modeling of colloid aggregation at surfaces
has a long history [3].
Recent studies of ballistic deposition on both linear and planar
substrates, mostly limited to lattice models, were largely focussed on
the properties of growing surfaces far away from the wall [4-5]. For
numerical investigations of amorphous assemblies of spheres produced by
means other then irreversible deposition, for modeling experimental
``powder'' properties, consult, e.g., [6]. Models of epitaxial growth
where particles fall on an ordered substrate (e.g., a lattice of
spheres) and further relax, for instance by rolling down to
the lowest lattice position after the
first contact, have been considered in the literature [7]
although most numerical simulations were limited to a lower-dimensional
case of deposition of circles.

The outline of the later sections is as follows. The model is defined,
and the numerical simulation details are given, in Section~2.
Section~3 is devoted to the presentation and discussion of our results
for the sphere-center density as a function of the distance from the
wall. Further discussion of the density properties near the wall, in
the framework of the chain-formation model, is the subject of
Section~4. Finally, Section~5 contains the presentation of our
results on the statistics of sphere contacts and summary.

\NP

\NI{\bf 2.~DEFINITIONS AND OUTLINE OF THE NUMERICAL METHOD}

\hphantom{AA}

Let the ball (sphere) diameter be $D$. We will measure all quantities in
dimensionless units. Thus, the substrate surface was square, of size
$\ell D \times \ell D$, with $\ell$ values of order 200 in
our simulations. Periodic boundary conditions were imposed in both
directions within the substrate to minimize finite-$\ell$ effects.
Each ball was positioned randomly above the substrate and then
``dropped'' vertically to stick at its first contact with
the substrate or one of
the earlier-dropped balls. The number of balls in each run was of order
$10^6$ and the results were averaged over many runs; specific numbers
and computer program details are given later.

We recorded the ball density and the statistics of contacts. The latter
will be described in Section~5. The ball-center density was measured
by binning the center coordinates in the histogram with bins at

$$ \left({1 \over 2}+{m \over K}\right)D \leq z <
 \left({1 \over 2}+{m+1 \over K}\right)D \; , \eqno(2.1) $$

\NI where $z$ is the distance of the ball center from the wall, while
$m=0,1,2,\ldots$ labels the bins. Let us denote the count in bin $m$,
averaged over many runs, as mentioned earlier, by $C_m$.

Note that the ball centers are located at distances $z \geq D/2$ away
 from the wall. However, we found that the count $C_0$ is much larger
than the counts $C_{m>0}\,$, due to the formation of a finite
surface-coverage density of balls in direct contact with the wall.
If we introduce the dimensionless variable

$$ h = {z\over D}-{1\over 2} \; , \eqno(2.2) $$

\NI then the total ball density can be formally written, for $h \geq
0$, as

$$ \big[ \rho (h) + \theta \delta (h) \big] D^{-3}
\; . \eqno(2.3) $$

\NI Here the wall-contact density contribution (per unit area) is
given by $\theta D^{-2}$, where $\theta$ can be estimated numerically
 from the relation

$$ \theta = \lim_{K\to \infty} \left( C_0 / \ell^2 \right) \; .
\eqno(2.4) $$

\NI The spatial ball-center density for $h>0$ is $\rho D^{-3}$
(per unit volume), where

$$ \rho (h) = \lim_{K\to \infty} \left( K C_{Kh} / \ell^2 \right)
\; . \eqno(2.5) $$

Results of our numerical simulations and their analysis will be
presented in later sections. In the remainder of this section we
outline some of the programming aspects of the simulation.
Readers interested in results only can skip now to Section 3.

In the deposition of the $n^{\rm th}$ sphere, we select its planar
coordinates $(X_n,Y_n)$ randomly and independently. However, the
vertical coordinate $z=Z_n$ must be determined by the first-contact
condition. Let $(X_m,Y_m,Z_m)$ denote the coordinates of the centers
of spheres deposited earlier, $m=1,2,\ldots, n-1$. We examine the
numbers $z_m$ defined by

$$ z_m  = Z_m + \sqrt{ D^2 - \left(X_n-X_m\right)^2 - \left(Y_n-Y_m
\right)^2} \; \; . \eqno(2.6) $$

\NI Most of these $(n-1)$ numbers will not be real. However, we only
keep those for which the argument of the square root is nonnegative
so that they are real (for positive arguments the positive root
value is taken). The resulting real-$z_m$ values and the number $D/2$
are compared and the {\sl largest} among all these numbers is the
required value $Z_n$.

A straightforward selection by maximization of (2.6),
among $(n-1)$ candidates $z_m$,
requires order $n$ computations for particle $n$
and hence order $N^2$ computations for depositing $N$ particles.
Our program actually spent only order $N$ computations
for $N$ particles. This was accomplished by splitting
the area of the substrate into sectors.
For each sector we maintained a list of particles deposited in it.
For each new particle,
our program checked particles only in a few neighboring sectors,
rather than all the previously deposited particles.
Moreover, the particles in the sector were ordered
according to their $z$-coordinates, and the
checking within each sector was restricted to a few top particles.

\NP

\NI{\bf 3.~RESULTS FOR THE DENSITY OF SPHERE CENTERS}

\hphantom{AA}

Our longest run took about 2 CPU weeks on a SUN SPARC workstation.
The substrate size was $\ell=200$. The number of balls dropped in each
run was $10^6$, and the results were averaged over $1005$ independent
runs. The bin size was $1/K=1/97$.
{}From the data collected in this run as
well as in other simulations (see further below), we estimate

$$ \theta = 0.318 \pm 0.001 \; . \eqno(3.1) $$

\NI Note that if the ball adhesion events
were allowed only on the substrate,
and not on other balls, then this system would be equivalent to the
Random Sequential Adsorption process of depositing disks on a plane
[1]. In the latter process, the surface coverage (fraction of area
covered) reaches the value $\sim 0.547$ at large times [8]. This
corresponds to the disk-center density $\sim 0.430$, in
units of $ D^{-2}$. The result
(3.1) is considerably lower indicating a significant screening
of the surface layer by balls deposited in higher layers.

As already mentioned the concept of a ``layer'' is used here loosely.
It turns out, however, that the deposit formed does show some tendency
to layering at least for distances up to about $h=5$. The density
of ball centers near the wall, as obtained in our longest run,
is shown in Figure~1. The fluctuations observed, suggesting the layered
structure, can be attributed to chain formation which will
be modeled in detail in the next section. The layering tendency is not
due to any ordering of a crystalline type. Examination of few snapshots
of the two-dimensional cross-sections parallel to the substrate
both near and away from the wall, for a run with the substrate size
$\ell=256$, suggests no crystalline regularity in the ball arrangement.

For distances $h^>_\sim 5$, the density oscillations fade away.
Instead, the density decreases monotonically. In earlier
numerical studies of ballistic deposit growth from point seeds
and on lattice substrates, mainly for low-dimensional, lattice
models [4,5,9,10], it was noted that the density away from the seed
or surface falls as a power law,

$$\rho(h) \simeq \rho(\infty ) + {\rm const}/h^p \; , \qquad
{\rm for} \qquad h \gg 1 \; , \eqno(3.2)$$

\NI with the exponent estimates spanning the range $p = 0.80 \pm 0.06$
for two-dimensional substrates;
see [4,5,11]. This behavior attracted much interest and it was attributed
[11] to the formation of large gaps due to surface roughening
which in turn arises from screening effects. As pointed out
earlier, the morphology of deposits far away from the substrate is
sensitive to the transport mechanism. The ballistic transport
yields screening too weak to cause formation of a ramified,
fractal structure. The power-law tail is thus the most profound
result of screening in ballistic deposits. In particular, the exponent
in (3.2) was related to the kinetic roughening exponents;
see [11] for further discussion.

Our longest-run data were recorded up to approximately
$h \simeq 20$. In order to check that the observed monotonic decrease
in density was not due to incomplete saturation,
and to make sure that finite-$\ell$ effects were
negligibly small, we made another long run. The number of spheres
was increased to $1.6 \cdot 10^6$, while the substrate size was
reduced to $\ell = 160$. The bin size was $1/K=1/41$, and the results
were averaged over 236 independent deposition runs. The data were
recorded up to approximately $h \simeq 50$. Comparison of the
results of the two long runs confirmed the observed monotonic
variation. The data from both runs are plotted
vs.~$h^{-0.8}$ in Figure~2. Despite the statistical noise, it can
be claimed that the power-law behavior is consistent with
the data and that the exponent in (3.2) is in the range $p=0.8 \pm
0.2$. The saturation density is

$$ \rho (\infty ) = 0.280 \pm 0.005 \; . \eqno(3.3) $$

\NI We also tried fit to the exponential decay law. However, the
power-law is clearly favored.

The asymptotic ``bulk'' density of sphere centers $\sim
0.28 D^{-3}$ corresponds to the packing fraction
(fraction of the volume filled up) of $\sim 15$\%. This is
considerably smaller than the packing fraction of random
assemblies of spheres [6] formed by ``bulk'' mechanisms with
relaxation. Indeed, in the latter processes the filled volume
fraction is typically over 60\%. Presently,
the most accurate estimate of the
packing fraction for the ball-deposition ballistic aggregation
model in three dimensions is $0.1465 \pm 0.0003$; see [4]. This range
corresponds to

$$ \rho (\infty ) = 0.2798 \pm 0.0006 \; , \eqno(3.4) $$

\NI in excellent agreement with our estimate (3.3).

\NP

\NI{\bf 4.~CHAIN MODEL OF THE DENSITY VARIATION NEAR A WALL}

\hphantom{AA}

Let us assume that the deposit growth can be described in some
approximate sense by the following chain-formation model.
The incoming balls attach to the end balls of chains of previously
deposited balls. Each chain starts with a ball on the substrate
and then the later-attached balls can be identified in a linear sequence.
Obviously, such a model can be at best approximate. Let us, however,
explore its implications.

Within an average, effective-field type prescription, we should be
able to assign the probability function $w(x)$
for the $m^{\rm th}$ ball in a given chain
to adhere with its center displaced
the distance $x$ (measured in units of $D$) from the center
of the ball $(m-1)$. This displacement is in the vertical,
$h$-direction. The function $w(x)$ is normalized,

$$ \int_{-\infty}^\infty w(x) dx =1 \; . \eqno(4.1)$$

\NI In fact, this function will be strongly localized in
$0 \leq x \leq 1$. The value $x=1$ corresponds to the head-on
deposition, while the value $x=0$ corresponds to the extreme
circumferential impact parameter equal the sphere diameter.

The simplest model is of course to assume that both the
surrounding-chain spheres and the preceding spheres in the same chain
do not interfere in any way with the deposition event. It is then
quite easy to check that on geometrical grounds alone
a sphere dropped with uniform probability distribution
over the cross-section of another sphere, will adhere with probability

$$ w(x)=2x \; \, \qquad {\rm for} \qquad 0\leq x\leq 1 \; , \eqno(4.2)
$$

\NI and $w(x)=0$ outside this range of the center displacement.

The first balls in the chains, those in direct contact with the wall,
contribute the term

$$\rho^{(1)} = \theta \delta (h)     \eqno(4.3)$$

\NI to the density (measured in units of $D^{-3}$); see (2.3).
The contribution of the balls $m>1$ can be calculated iteratively,

$$ \rho^{(m)}(h) = \int_{-\infty}^\infty \rho^{(m-1)} (h-x) w(x) dx
\; , \eqno(4.4) $$

\NI where the total spatial density in (2.3) is then given by

$$ \rho (h) = \sum_{m=2}^\infty \rho^{(m)} (h) \; . \eqno(4.5) $$

Already the simplest model distribution (4.2), with $\theta = 0.318$
taken from the numerical estimation, see (3.1), gives the density
function which has many semi-quantitative similarities with the
measured density near the wall. This function was evaluated numerically
and plotted in Figure~3. It should be compared with Figure~1; both
figures were plotted with the same axes ranges. Note that the
resulting density function $\rho (h)$ has a discontinuity at $h=1$, as
well as a discontinuous $(h-1)$-order derivative at each
integer value $h>1$.

The chain-model ideas in deposition, in a somewhat different context,
have already been used in the
literature; see, e.g., [10] where off-lattice
ballistic deposition of circles on a seed was studied. While it is
tempting to modify the simplest distribution (4.2) to improve the
quantitative consistency with the measured density, the chain model
should not, in fact, be taken too seriously. Firstly, at short distances
it is not quite clear if the actual density will have true
derivative-discontinuities near integer $h$, or just sharp but rounded
anomalies. This is difficult to decide from the numerical data
available. Within the chain model, one would then round up the $w(x)$
function near $x=0$ and $x=1$.

At large distances, the chain-model prediction for the density is

$$ \rho(\infty) = \theta / x_1 \; , \eqno(4.6) $$

\NI where $x_1$ is the first-moment displacement,

$$ x_1 = \int_{-\infty}^\infty x w(x) dx \; . \eqno(4.7) $$

\NI However, our numerical estimates of $\theta$ and $\rho ( \infty ) $
suggest $x_1 \simeq 1.14 > 1$, see
(3.1) and (3.3). Thus, the ``realistic'' distribution $w(x)$ must
``protrude'' past $x=1$ to fit the large-$h$ data! (Note that the
large-$h$ limiting value in Figure~3 is 0.477.)

Further difficulty is suggested by the power-law tail in the density,
(3.2). The Fourier-space considerations not detailed here [note that
the convolution (4.4) becomes a product in the Fourier space],
suggest that this tail implies long-range tails in the distribution
$w(x)$ as well. All these observations make it actually quite
difficult to propose a plausible few-parameter
form for $w(x)$ to fit all the
features of the observed density variation within the chain model.

We therefore adopt the point of view that the chain model can be used
(a) to describe the behavior of the saturated-deposit
density within few diameters from the
wall semi-quantitatively; (b) to suggest some general features of the
density to be checked by future numerical simulations such as possible
discontinuities in $\rho(h)$ and its derivatives; (c) in
time-dependent deposition modeling for short times when the average
coverage is within few diameters.

Note that relations (4.3)-(4.5) for the density are recursive
\UN{along the chains} which are anyway approximate objects. In a series
of papers [12], Savit and coworkers considered approximate recursive
relations \UN{in time}, for particle adhesion probability distribution
within the growing kinetically-roughened interface. Their conclusion
was that in many models quasiperiodic irrational-frequency
density fluctuations should be present, with density fluctuating on
scales larger than the underlying particle-size ``clock'' length, 1
in our reduced units (2.2). It should be emphasized that fluctuations
in density and other quantities (see Section 5) found in our work are
\UN{not} the ``bulk'' effect described in [12]. In fact, what we are
observing is the ``clock'' particle-size fluctuations with underlying
periodicity 1. These fluctuations fade away for distances over 5-6
particle diameters from the wall. Therefore, the irrational-frequency
fluctuations [12] cannot be present in the model considered here.
Indeed, far from the wall even the basic particle-size periodicity
is not preserved; the deposit is truly amorphous. However, ``bulk''
density fluctuations can be present in lattice ballistic deposition
[13], as well as in off-lattice models where the underlying
discrete structure is preserved far from the wall.
This can be achived by having lattice substrate and relaxation
mechanisms for particles to align themselves with the deposit structure on
adhesion, or by having particle shapes, e.g., oriented cubes, which
force layer structure in the bulk.

\NP

\NI{\bf 5.~CONTACT STATISTICS}

\hphantom{AA}

One of the conclusions of Sections~2-3 was that the deposit formed by
ballistic transport and irreversible sticking is quite different from
the uniform assembly of randomly packed spheres [6]. We found that
the density is much
lower, and the structure has a preferred orientation and tendency to
layering at least within the first few diameters from the wall.
It is well established that the average number of spheres in contact
with each given sphere in the randomly packed structure is near 6;
see [6].

We collected the statistical data on the number of contacts per ball,
binned similarly to the density statistics described in Section~2.
We only found contacts with 1, 2, 3, 4, 5, and in very few instances,
with 6 balls (the count included wall contacts as well). Let us denote
by $f_i (h)$ the fraction of balls with $i=1,2,3,4,5,6$ contacts, at
the (dimensionless) distance $h$ from the wall; see (2.2). The average
number of contacts is given by

$$ \sum_{i=1}^6 i f_i (h) \; , \qquad {\rm where} \qquad
 \sum_{i=1}^6 f_i (h) =1  \; . \eqno(5.1) $$

\NI This function, as estimated from our longest run (Section~2),
is plotted in Figure~4 for $h>0$. It shows general pattern of behavior
similar to the center density: there are oscillations and
(possibly rounded) discontinuities in the function or
its derivatives near integer values $h=1,2,\ldots$.
Although not shown in the figure, the value $\simeq 1.900$
found at $h=0$ suggests also the discontinuity as $h \to 0$.

The large-$h$ behavior of the average number of contacts can be
represented by a power-law relation similar to (3.2). The data from
the two long runs are quite straight when plotted vs.~$h^{-0.8}$
\ \ (cf.~Figure~2), suggesting that the same exponent
$p \simeq 0.8$ applies here.
The limiting value as $h \to \infty$ was estimated as $2.000 \pm
0.005$. Early numerical studies [3] of ballistic deposition also
yielded ``bulk'' contact numbers near 2. In fact, one can prove that
this limiting value is exactly 2,
provided the average number of contacts approaches a constant
value as $h\to \infty$, i.e., the fluctuations are damped,
as is indeed suggested by our data.
Note that this ``coordination number'' is much lower than
the typical ``powder'' values $\sim 6$ quoted earlier.

The fractions of contacts $f_i (h>0)$
for $i=1,2,3$ are plotted in Figure~5.
The remaining contacts, $i=4,5,6$, amounted to at most 2\%, typically
less. Thus, Figure~5 summarizes the main contact-statistics properties.
Each fraction has (possibly rounded)
discontinuities similar to the density and average
number of contacts. There is also the discontinuity at $h=0$ where the
values, not shown in the figure, were $\simeq 0.286, 0.541, 0.161$, for
$i=1,2,3$, respectively.

Note that the simplest chain model would correspond to $f_2=1$ and
the number of contacts 2. In actuality, however, far from
the wall only about 53\% of the balls have two contacts. About 25\%
of the balls have only one contact which indicates that they were
screened from the incoming flux of balls by the surrounding structure,
thus becoming ``dangling ends.''
On the other hand, the remaining 22\% of the balls have 3 (about 20\%)
or more (about 2\%) contacts thus forming branching points in the
structure. Qualitatively, these conclusions were also confirmed
by visual examination of a ``snapshot''
of the ball configuration in the plane perpendicular to the substrate.

The above percentages were far from the wall. Figure~5 suggests that
both screening and branching change abruptly near integer $h$
values within the first few diameters from the wall, which correlates
with the density fluctuations, although we are not aware of any
theoretical modeling of such properties.

In summary, we reported the first detailed investigation of the
morphology of ballistic deposits near walls. The observed structure is
disordered but an approximate notion of ``layers'' can be used near the
wall. Deposits formed by irreversible sticking are much sparser
than the relaxed, powder-type structures. On the average, each particle
is in contact with two others particles. The chain model in its
simplest form accounted only for the near-wall layering while the more
``bulk'' properties seem to be governed by competition of screening
and branching.

The authors are indebted to Dr.~J.~Krug for helpful comments and
suggestions.
This research was partially supported by the
Science and Engineering Research Council (UK)
under grant number GR/G02741.
One of the authors (V.P.) also wishes to acknowledge
the award of a Guest Research Fellowship at Oxford
from the Royal Society.

\NP

\centerline{\bf REFERENCES}

\hphantom{AA}

\item{1.} For comprehensive review and literature list, consult
M.C.~Bartelt and V.~Privman, Int.~J.~Modern Phys.~B{\bf 5}, 2883 (1991).

\item{2.} For recent results and further literature, consult, e.g.,
N. Ryde, N. Kallay and E. Matijevi\'c,
J. Chem. Soc. {\sl Faraday Trans.\/} {\bf 87}, 1377 (1991);
M. Elimelech and C.R. O'Melia, Environ. Sci. Technol.
{\bf 24}, 1528 (1990). Additional literature citations can be found in
[1].

\item{3.} M.J. Vold, J. Colloid Sci. {\bf 14}, 168 (1959);
J. Phys. Chem. {\bf 63}, 1608 (1959) and {\bf 64}, 1616 (1960).

\item{4.} R. Jullien and P. Meakin, Europhys. Lett.
{\bf 4}, 1385 (1987).

\item{5.} P. Meakin and R. Jullien, Phys. Rev. A{\bf 41}, 983 (1990);
J. Krug and P. Meakin, J. Phys. A{\bf 23}, L987 (1990);
R. Baiod, D. Kessler, P. Ramanlal, L. Sander
and R. Savit, Phys. Rev. A{\bf 38}, 3672 (1988).

\item{6.} J.G. Berryman, Phys. Rev. A{\bf 27}, 1053 (1983);
W.S. Jordey and E.M. Tory, Phys. Rev. A{\bf 32}, 2347 (1985).

\item{7.} These studies are reviewed, e.g., in: J.W. Evans, Vacuum {\bf
41}, 479 (1990).

\item{8.} E.L. Hinrichsen, J. Feder and T. J{\o}ssang, J. Stat. Phys.
{\bf 44}, 793 (1986).

\item{9.} D. Bensimon, B. Shraiman and S. Liang, Phys. Lett.
{\bf 102}A, 238 (1984); S. Liang and L.P. Kadanoff, Phys. Rev.
A{\bf 31}, 2628 (1985).

\item{10.} P.S. Joag, A.V. Limaye and R.E. Amritkar, Phys. Rev.
A{\bf 36}, 3395 (1987).

\item{11.} J. Krug, J. Phys. A{\bf 22}, L769 (1989).

\item{12.} R. Baiod, Z. Cheng and R. Savit, Phys. Rev. B{\bf 34},
7764 (1986); Z. Cheng and R. Savit, J. Phys. A{\bf 19}, L973 (1986);
R. Savit and R.K.P. Zia, J. Phys. A{\bf 20}, L987 (1987);
Z. Cheng, R. Baiod and R. Savit, Phys. Rev. A{\bf 35}, 313 (1987).

\item{13.} Z. Cheng, L. Jacobs, D. Kessler and R. Savit,
J. Phys. A{\bf 20}, L1095 (1987).

\NP

\centerline{\bf FIGURE CAPTIONS}

\hphantom{AA}

\NI\hang {\bf Fig.~1.}~$\;$Spatial density of ball centers evaluated
numerically in the long run with bin size $1/K=1/97$, see Section 3. The
histogram counts were multiplied by $K/\ell^2$
according to (2.5), to yield the
density estimates which were plotted at the
$h$ values corresponding to the
centers of the bins; cf.~(2.1)-(2.2). The lines connecting three points
near $h=1$ were drawn to guide the eye.

\NI\hang {\bf Fig.~2.}~$\;$Density of ball centers vs.~$1/h^{0.8}$.
The data shown are from the two long runs as described in Section~3.

\NI\hang {\bf Fig.~3.}~$\;$Density of ball centers calculated with the
simplest chain-model probability function (4.2).

\NI\hang {\bf Fig.~4.}~$\;$Average number of contacts for spheres with
centers at $h>0$; see (5.1).

\NI\hang {\bf Fig.~5.}~$\;$The fraction of spheres with 1, 2, and 3
contacts (marked by $f_{i=1,2,3}$), for $h>0$.
The remaining contact fractions, for 4, 5, and 6 contacts, sum up
to at most $\sim 0.02$ and are not shown.

\hphantom{A}

\NI {\bf To get the preprint with the figures write to \NL
PRIVMAN@CRAFT.CAMP.CLARKSON.EDU}

\bye